\input harvmac
\input epsf

\noblackbox


\def\bfone{\relax{\rm 1\kern-.35em 1}}
\def\inbar{\vrule height1.5ex width.4pt depth0pt}

\def\IC{\relax\,\hbox{$\inbar\kern-.3em{\rm C}$}}
\def\ID{\relax{\rm I\kern-.18em D}}
\def\IF{\relax{\rm I\kern-.18em F}}
\def\IH{\relax{\rm I\kern-.18em H}}
\def\II{\relax{\rm I\kern-.17em I}}
\def\IN{\relax{\rm I\kern-.18em N}}
\def\IP{\relax{\rm I\kern-.18em P}}
\def\IQ{\relax\,\hbox{$\inbar\kern-.3em{\rm Q}$}}
\def\us#1{\underline{#1}}
\def\IR{\relax{\rm I\kern-.18em R}}
\font\cmss=cmss10 \font\cmsss=cmss10 at 7pt
\def\ZZ{\relax\ifmmode\mathchoice
{\hbox{\cmss Z\kern-.4em Z}}{\hbox{\cmss Z\kern-.4em Z}}
{\lower.9pt\hbox{\cmsss Z\kern-.4em Z}}
{\lower1.2pt\hbox{\cmsss Z\kern-.4em Z}}\else{\cmss Z\kern-.4em
Z}\fi}
\def\a{\alpha}  
\def\e{\epsilon} 
 \def\l{\lambda}
 \def\s{\sigma}
\def\th{\theta}

\def\cL{{\cal L}} 
\def\cN{{\cal N}}

\def\nup#1({Nucl.\ Phys.\ $\us {B#1}$\ (}
\def\plt#1({Phys.\ Lett.\ $\us  {B#1}$\ (}
\def\cmp#1({Comm.\ Math.\ Phys.\ $\us  {#1}$\ (}
\def\prp#1({Phys.\ Rep.\ $\us  {#1}$\ (}
\def\prl#1({Phys.\ Rev.\ Lett.\ $\us  {#1}$\ (}
\def\prv#1({Phys.\ Rev.\ $\us  {#1}$\ (}
\def\mpl#1({Mod.\ Phys.\ Let.\ $\us  {A#1}$\ (}
\def\ijmp#1({Int.\ J.\ Mod.\ Phys.\ $\us{A#1}$\ (}
\def\jag#1({Jour.\ Alg.\ Geom.\ $\us {#1}$\ (}
\def\tit#1|{{\it #1},\ }

\def\Coe#1.#2.{{#1\over #2}}
\def\coeff#1#2{\relax{\textstyle {#1 \over #2}}\displaystyle}
\def\coe#1.#2.{\relax{\textstyle {#1 \over #2}}\displaystyle}
\def\half{{1 \over 2}}
\def\shalf{\relax{\textstyle {1 \over 2}}\displaystyle}

\def\del{\partial}

\def\Fh#1{{{}_3F_2\left(\coeff{1}{4},\coeff{3}{4},\coeff{3}{4}
;\coeff{1}{2},\coeff{5}{4};#1\right)}}
\def\Gh#1{{{}_3F_2\left(\coeff{3}{4},\coeff{5}{4},\coeff{5}{4}
;\coeff{3}{2},\coeff{7}{4};#1\right)}}
\def\Fg#1{{{}_2F_1\left(-\coeff{1}{4},\coeff{3}{4};\shalf;#1\right)}}
\def\Gg#1{{{}_2F_1\left(\coeff{1}{4},\coeff{5}{4};\coeff{3}{2};#1\right)}}

%
\lref\JMalda{J. Maldacena, {\it The large N limit of superconformal
field theories and supergravity}, hep-th/9711200.}
\lref\MDJR{M.J.~Duff and J.~Rahmfeld  \nup{481} (1996) 332, 
hep-th/9605085.}
\lref\KStelle{K.S.~Stelle  {\it BPS Branes in Supergravity,} 
in Proceedings of the ICTP Summer School in High-energy Physics 
and Cosmology, Trieste, Italy, 10 Jun -- 26 Jul 1996 and 2 Jun -- 
11 Jul 1997, hep-th/9803116.}
\lref\JMaldb{J. Maldacena, {\it Wilson loops in large N field 
theories}, HUTP-98/A014, hep-th/9803002.}
\lref\RY{S.-J. Rey and J.~Yee, {\it Macroscopic strings as heavy quarks in 
large {N} gauge theory and anti-de Sitter supergravity,} hep-th/9803001.}
\lref\RTY{S.-J. Rey, S.~Theisen and J.~Yee, {\it Wilson-Polyakov Loop at 
Finite Temperature in Large $N$ Gauge Theory and
     Anti-de Sitter Supergravity,} hep-th/9803135.}
\lref\BISY{A.~Brandhuber, N.~Itzhaki, J.~Sonnenschein and S.~Yankielowicz,
{\it Wilson Loops in the Large $N$ Limit at Finite Temperature}, 
hep-th/9803137}
\lref\BISYII{A.~Brandhuber, N.~Itzhaki, J.~Sonnenschein and S.~Yankielowicz,
{\it Wilson Loops, Confinement, and Phase Transitions in Large $N$ Gauge 
Theories from Supergravity}, hep-th/9803263.}
\lref\ML{M.~Li, {\it 't Hooft Vortices on D-branes}, hep-th/9803252;
 {\it 't Hooft vortices and phases of large $N$ gauge theory},
hep-th/9804175.}
\lref\DP{ U.~Danielsson and  A.~Polychronakos, {\it Quarks, monopoles and 
dyons at large $N$,} hep-th/9804141.}
\lref\JMin{J.~Minahan, {\it Quark-Monopole Potentials in Large $N$ Super 
Yang-Mills}, hep-th/9803111.}
\lref\LanLif{L.D.~Landau and E.M.~Lifshitz, {\it Mechanics,}
Third Edition, Pergamon Press, (1973).}
\lref\Wittherm{E.~Witten, {Anti-de Sitter Space, Thermal Phase Transition, 
And Confinement In Gauge Theories}, hep-th/9803131.}

%
%
\Title{\vbox{
\hbox{USC-98/006}
\hbox{\tt hep-th/9805104}
}}{\vbox{\centerline{\hbox{Quark Potentials in the Higgs 
Phase of Large $N$}}
\vskip 8 pt
\centerline{ \hbox{ Supersymmetric Yang-Mills 
Theories}}}}
\centerline{J.~A.~Minahan and N.~P.~Warner}
\bigskip
\centerline{{\it Physics Department, U.S.C., University Park,
Los Angeles, CA 90089-0484, USA}}
\bigskip

We compute, in the large $N$ limit, the quark potential for 
${\cal N}=4$ supersymmetric $SU(N)$  Yang-Mills theory broken 
to $SU(N_1) \times SU(N_2)$.   At short distances the quarks see only
the unbroken gauge symmetry and have an attractive potential that 
falls off as $1/L$.  At longer distances the interquark interaction
is sensitive to the symmetry breaking, and other QCD states appear.
These states correspond to combinations of the quark-antiquark pair
with  some  number of $W$-particles.  If there is one or more 
$W$-particles then this  state is  unstable because of the
coulomb interaction between the $W$-particles and between the
$W$'s and the quarks.  As
$L$ is decreased the $W$-particles delocalize and these coulomb branches
merge onto a branch with a linear potential.  The quarks on this branch
see the unbroken gauge group, but the flux tube is unstable to the
production of $W$-particles.

\vskip .3in

\Date{\sl {May, 1998}}
\vfil
\eject
\parskip=4pt plus 15pt minus 1pt
\baselineskip=15pt plus 2pt minus 1pt
%
\newsec{Introduction}

The conjectured link \JMalda\ between large $N$ $\cN=4$ 
supersymmetric Yang-Mills and
ten dimensional supergravity on $AdS_5\times S_5$ has led to a number
of interesting predictions.  In particular one has been able to compute
the large $N$ behavior for the attractive potential between  heavy
quarks and antiquarks, both at zero temperature \refs{\RY,\JMaldb}, and
finite temperature \refs{\Wittherm,\RTY,\BISY}, and for a quark and a 
monopole \JMin, collections of dyons, monopoles and quarks\DP, 
and flux tubes \refs{\ML}.
In these calculations, one looks for static string configurations that
correspond to a separation $L$ and potential energy $E$
between the heavy particles.  

For a quark-antiquark pair at zero temperature there is 
a unique string configuration for a given separation $L$ \refs{\RY,\JMaldb}.  
At finite 
temperature, one finds that there are some values of $L$ where two solutions
exist \refs{\RTY,\BISY}.  
The branch with the lower energy is the coulomb branch, where at
short distances, the quark and anti-quark experience an attractive
coulombic force.  The other branch is a ``confining'' branch, in that
the force is linear in $L$, at least for small $L$.  The
physical meaning of this branch is unclear.

There are other systems where more than one branch exists.  
As we will see below, an $SU(N)$ gauge theory broken to 
$\prod_i SU(N_i)$ with $gN_i$ large 
will have many branches.  In fact as $L$ diverges, the number of
branches diverges.  In this paper we will study the breaking of
$SU(N)$ to $SU(N/2)\times SU(N/2)$.   In this situation some of
the  string configurations can be solved by quadratures,
or exactly in various limits, and the
qualitative features can be generalized to other breakings. 

The physics of these extra branches can be deduced by looking at the
energy density along the static string configuration, and also by 
taking the quark separation to infinity.  In particular we find 
static string configurations that correspond to 
states consisting of a quark, an antiquark 
and $n$ massive $W$ particles, with the latter 
transforming in the $(N/2,\overline N/2)$. The energy density along
the string concentrates appropriately: there are two (infinitely) massive
quarks at the ends and $n$ equal and evenly spaced mass concentrations
in between.  In the limit of infinite separation the total
energy of the system becomes the sum of the BPS quark masses plus the 
BPS masses of $n$  $W$-particles, while
the energy density reduces to that of the quarks and $n$ evenly 
spaced $\delta$-functions for the $W$-particles.  The static
string configurations in this limit also decompose into a 
concatenation of such configurations for a quark, an antiquark and  
$n$ intermediate $W$'s.   As the quark separation for the
$n$ $W$-particle configuration is decreased the branch
becomes unstable to small perturbations
 because of the attractive coulomb potential between the $W$ 
particles and the quarks.  As $L$ is decreased further the $W$-particles 
delocalize and the unstable branches
merge onto a branch with a  potential that is linear for large enough $L$.  

Turning this around, for large enough quark separation there is
an unstable flux tube configuration that has a linear potential between 
the quarks.  This is an unbroken gauge configuration in that the 
quark and anti-quark are both in the fundamental of $SU(N)$.  
The flux tube can then destabilize by popping out $W$-particles 
that are then subject to the coulomb instability.  
Eventually, a stable configuration is reached where we have 
a quark and an anti-quark in $(N/2,1)$ or $(1,N/2)$
representations of the broken group.  Finally, as $L$ is decreased even
further, these stable branches merge onto the unbroken gauge branch.  
For $L$ below this critical value this is the only string solution.

In section 2 we discuss some general features for static string 
configurations and discuss how the study of such configurations
reduces to the study of particle motion in a potential, $V$.
We use Maupertuis' principle to argue that the force between the 
quarks is always attractive and has a simple relation to the classical 
particle's energy.  We then specialize to geodesics that correspond 
to an $SU(N)$ gauge theory broken to $SU(N/2)\times  SU(N/2)$, for 
which the potential, $V$, is  bipolar and $\ZZ_2$ symmetric.
In section 3 we consider string trajectories that lie on the symmetry
axis.  We show that there is a branch of trajectories  that runs up a 
saddle and which is coulombic for small separations and confining for 
large.   We also show that there are infinitely many branches that 
merge onto this confining branch.    All but the lowest energy 
branches are unstable to small perturbations and  eventually become 
coulombic as the separation is increased.  In section 4 we consider 
trajectories that are off the symmetry axis.  In this case, the quark 
and anti-quark have unequal masses.  The branches split into groups of 
three, except for the lowest energy branch which is isolated.  
The higher energy branches have a minimum quark separation.
Pushing the quarks inside of this separation leads to a finite release
of energy.  Section 5 contains some final comments.

\newsec{Metrics, Geodesics and the Quark Potential }

\subsec{Generalities}

The general form of the metric for a configuration of
parallel, static  D$3$-branes in ten dimensions is:
\eqn\genmetric{ds^2 ~=~ H^{-1/2}~dz^\mu dz^\nu \eta_{\mu\nu} 
{} ~+~ H^{1/2}~dy^mdy^m \ ,}
where $\mu = 0, \dots,3$ and $m = 4,\dots,9$, and $H$ is
a function of the $y^m$ alone.

Our purpose is to use the ideas first mooted in
\JMalda, and further developed in \refs{\RY{--}\ML}, to
study the quark-antiquark potential.
This means that we consider static string configurations that
begin and end on a single brane at a very
large (and ultimately infinite) distance, $\vec y = \vec y_{max}$,
from the rest of the branes in the configuration defined by
$H$.  Finding such static string
configurations is equivalent to finding geodesics in the
spatial $9$-metric:
\eqn\effmetric{ds^2 ~=~ H(y)^{-1}~dz^\alpha dz^\alpha  ~+~ 
dy^mdy^m \ ,}
where $z^\alpha$, $\alpha = 1,2,3$, runs over the spatial
coordinates of the D$3$-brane.  Because of the symmetry
in the $z^\alpha$, we can choose coordinates so that the
geodesic runs in the $z^1 = z$ direction, with $z^2 = z^3 =0$.
Let $\lambda$ be an affine parameter for geodesics in the metric
\effmetric, then the translational symmetry implies that
${dz \over d \lambda} = const~H(y)$.  We will normalize the
affine parameter by taking the constant equal to one, 
and hence
\eqn\conslaw{{dz \over d \lambda} ~=~ H(y).}
Using this conservation law in the geodesic equations for 
$y^m(\lambda)$ yields:
\eqn\yaccel{{d^2 y^m \over d \lambda^2} ~+~ \half
{\del H \over \del y^m}  ~=~ 0\ .}
We are thus studying the motion of a particle in the
(repulsive) potential $V \equiv \half H$.  We will denote
the classical energy of this particle by:
\eqn\parten{\mu  ~=~ \half \Big({d y^m 
\over d\lambda}\Big)^2 ~+~ \half H (y)\ .}
The energy
of the quark-antiquark pair is given by
\eqn\qenergy{E~=~ {1\over 2 \pi \alpha'}~ \int \sqrt{\Big({d y^m 
\over d\lambda}\Big)^2 ~+~ H (y)}~~d \lambda ~=~ 
{1\over 2 \pi \alpha'}~\sqrt{2 \mu}~\lambda\ ,}
and the three-dimensional separation, $L$, of the 
quark-antiquark pair is given by:
\eqn\fluxL{L ~=~ \int H (y(\lambda))~d \lambda \ .}
The energy density along the string can be readily deduced from
\qenergy\ and \fluxL, with the density $\rho$ given by
\eqn\endens{ \rho~=~{1\over2\pi\a'}~{\sqrt{2\mu}\over H(y)} \ .}

We ultimately wish to study $E$ as a function of $L$,
and seek out coulombic and confining behavior.  One can
gain some very useful insight into $E(L)$ by applying
some of the variants of variational principles associated
with particle motion, most particularly, Maupertuis' principle
\LanLif.  Consider a general variation of the
action $S({d \vec q \over dt}, \vec q)$ about a solution
of the Euler-Lagrange equations, but for which we let the 
endpoints and the time for the path vary:
\eqn\genvarS{\delta S ~=~ (\vec p_f \cdot \delta  \vec q_f - H_f
\delta t_f) ~-~ (\vec p_i \cdot \delta  \vec q_i - H_i
\delta t_i) \ ,}
where the subscripts $i$ and $f$ denote initial and final
quantities, while $\vec p$ is the particle momentum,
and $H$ is the Hamiltonian.  We will  consider variations
for which $\vec p \cdot \delta  \vec q = 0$, and in partcular
variations in which the endpoints do not move.  Since the system 
has time translation invariance with a conserved
energy, $\mu$, we have $\delta S = - \mu \delta t$.
For such a system, it immediately follows that the
particle path is obtained from the variational
principle $\delta S_0 = 0$, where $S_0$ is the 
{\it abbreviated action}:
\eqn\abbrevS{S_0 ~=~ S + \mu(t_f - t_i) ~=~ \int \vec p 
\cdot d \vec q \ .}
In doing this variation one considers all paths with
a fixed energy, $\mu$, and one must express the integrand
of $S_0$ in terms of $\mu, \vec q$ and $d \vec q$ only.
For example, for a particle moving in a potential $V$, one has:
\eqn\Sopart{S_0 ~=~ \int \sqrt{2(\mu - V) d \vec q \cdot 
d \vec q } \ .}
Consider now, the variation of the path with respect to the 
energy, $\mu$.  The time taken along the path will change
in response, and so we have from \abbrevS:
\eqn\varSo{\delta S_0 ~=~   \delta S ~+~ (t_f - t_i) \delta 
\mu ~+~ \mu \delta t  ~=~   (t_f - t_i) \delta \mu\ , }
and so ${\partial S_0 \over \partial \mu}~=~ (t_f - t_i)$.

Returning to the original problem, we want to consider how the
quark energy, $E$ and separation, $L$, change as we vary the
geodesic, and in particular, vary the classical particle energy,
$\mu$.   From \qenergy, \fluxL\ and \varSo, one has
\eqn\varE{\eqalign{\delta E ~=~  &{1\over \pi \alpha' 
\sqrt{2 \mu}}~ \big(~\mu~ \delta \lambda   ~+~ \coeff{1}{2} 
\lambda~  \delta \mu ~\big) \ , \cr
\delta L ~=~  &2 \delta \int V ~=~  \delta ( S_0 - 2 S) ~=~ 
\lambda \delta \mu ~+~ 2 \mu \delta \lambda  \ . }}
{}From this one immediately sees that
\eqn\endres{\delta E ~=~  {1\over 2 \pi \alpha' 
\sqrt{2 \mu}}~ \delta L \ , \qquad {\rm or} \qquad 
{d E \over d L} ~=~    {1\over 2 \pi \alpha' 
\sqrt{2 \mu}} \ .}
Thus the force between a quark and anti-quark is determined
by the energy of the classical particle.

There are several important consequences of this.  First,
the force is always attractive:  something that is obvious
from QCD, but not at all obvious from the string perspective.
In particular, if one considers  $E$ and $L$ as a function
of some other parameter, as one does in the thermal case 
\refs{\RTY,\BISY},
one sees that $E$ and $L$ must reach a minimum or maximum at the
same point.  Hence if one of these functions ``turns over''
then so must the other.  Such simultaneous turning points of 
$E$ and $L$ are crucial to the cusp-like behavior $E(L)$ at
bound state transitions.  This will be discussed in more detail in 
section 4.

On a more fundamental level, to compute the quark potential
one considers a family of geodesics going out to the brane
at infinity.  A confining family is thus characterized by 
a family of such geodesics for which the classical particle 
energy is constant.  One can directly verify this in a special 
case: the situation 
where the family consists of a particle that is approaching
a local maximum, or saddle point from below, and turning 
around and reversing its course just before the maximum, or 
saddle.  Suppose the potential at the maximum or saddle
is $\mu_0$, and consider a particle coming in with energy 
$\mu_0 - \epsilon$.  For the particle to approach the
maximum, the time taken diverges as $t \sim - log(\epsilon)$,
and so the inter-quark energy $E$, and length $L$, also both
diverge as $- log(\epsilon)$, which means that $E \sim L$.

It is also interesting to consider the behavior of $E$ and
$L$ for particles that approach, or are released from a point
where the potential is diverging as $V(q) \sim q^{-r}$.
One can easily see that $L \sim \mu^{r-2 \over 2r}$ and
$E  \sim \mu^{-{1 \over r}}$ as one approaches the 
singularity\foot{One must, of course, subtract off the quark
mass to get this fall off in the energy.}.  One thus finds
$E \sim L^{-{2 \over r- 2}}$, which is consistent with \endres,
and shows Coulombic behavior only for $r=4$.   We can also find the
energy density along $z$, where we see that the density is 
$\rho\sim \mu^{-1/2}$ for the part of the string near the peak and 
$\rho\sim \mu^{1/2}$ for the part that is away from the peak. 
Hence, very energetic particles correspond to string configurations where
the energy density is highly clumped.

\subsec{Multi-centered D$3$-branes}

In \JMalda\ a single group of $N$ D$3$-branes was considered,
and the function $H$ was given by:
\eqn\oneBS{ H~=~ 1 ~+~ {4 \pi g N \alpha'^2 \over r^4} \ , 
\qquad r ~\equiv~  \sqrt{y^m y^m} \ .}
In the large $N$ limit, one then has the anti-de Sitter metric
\eqn\oneBmetric{ds^2 ~=~ {U^2 \over R^2}~dz^\mu dz^\nu 
\eta_{\mu\nu}  ~+~ R^2 {d U^2 \over U^2} ~+~ R^2 
d \Omega_5^2 \ ,}
where $U = r/\alpha'$, and $R = (4 \pi g N)^{1/4}$.  It was 
also pointed out in \JMalda\ that one could equally well
consider separated groups of D$3$-branes, and that for
large $N$ and large $U$ these would also limit to
anti-de Sitter space.

The multi-centered D$3$-brane solution is then given by
taking \refs{\MDJR,\KStelle}:
\eqn\Hmulti{H(y) ~=~ 1 ~+~ \sum_a ~{4 \pi g N_a \alpha'^2 
\over |\vec y ~-~  \vec y_\alpha|^4} \ , }
where $N_a$ is the number of branes located at the point
$\vec y_a$.   As in \JMalda, we consider the limit of the metric  
in which $\alpha' \to 0$ with $r/\alpha'$ finite, and in which
all of the $N_a$ are large  (but $N_a/N_b$ is finite).  
This means that we may drop the $1$ in $\Hmulti$.   The IIB 
superstring in this background should then describe a Higgs phase 
of an $N=4$ supersymmetric $SU(N)$ gauge theory, with 
$N = \sum_a N_a$, where the gauge group 
has been broken to $\prod_a SU(N_a)$.  The Higgs vevs are
represented by the separations of the groups of D$3$-branes.
In particular, at large $r = \sqrt{y^m y^m}$, the function 
$H$ in \Hmulti\ takes the form \oneBS\ with $N = \sum_a N_a$, 
and thus at large distances, and for large $N_a$, 
the multi-centered solution  limits to the  anti-de 
Sitter solution corresponding to $SU(N)$.

Our major focus will be the breaking of $SU(N)$ 
to $SU(M) \times SU(M)$, $M = N/2$, and upon geodesics in 
the corresponding  double-centered
metric. By rotating the coordinates, we can take the
potential  $V = {1 \over 2} H$ to be:
\eqn\doublepot{ V(x,y) ~=~ {\alpha'^2 \over 4}~\bigg[~{R^4 \over 
((x - \alpha' \phi)^2 +  y^2)^2} ~+~ {R^4 \over ((x+
\alpha' \phi)^2 + y^2)^2} ~\bigg] \ ,}
where we have set $y^1 = x, y^2 = y$.
To make contact with the conventions of other authors 
\refs{\JMalda{--}\ML}, in whose work $\alpha'$ has been scaled
away and in which the variables $x$ and $y$ have dimensions of 
mass, one makes the rescaling $x \to \alpha' x$, $y \to \alpha' y$
and $\lambda  \to \alpha'^2 \lambda$.

We will analyze the geodesic structure extensively in the
next section, but here we note some general features.
First, for large $r = \sqrt{x^2 + y^2}$, one has $V = \alpha'2  
R^4/(2r^4)$,
and provided the geodesic stays out at large values of $r$,
($r >> \alpha' \phi$)
the properties of such a geodesic will not differ significantly
from those discussed in \refs{\RY,\JMaldb}.  In the language of gauge theory
this means that the symmetry breaking mass scale of $SU(N)$ 
to $SU(M) \times SU(M)$, which is proportional to
$\phi$, is much less than the energy scale of the
quark-antiquark interaction ($1/L$), and so the interaction
behaves as it would in the unbroken $SU(N)$ phase.

At the other extreme, there are two geodesics that represent
single quarks:  That is, the geodesics
start on the single D$3$-brane and terminate on one or
other of the groups of $M$ D$3$-branes.  They traverse no distance
in the $z$-direction (the constant in the conserved quantity
is zero) and run with infinite classical particle speed.  
Such geodesics represent quarks in the $(M,1)$ or $(1,M)$ of
$SU(M) \times SU(M)$ and the quark mass is simply the distance in 
$\vec y$-space from $\vec y_{max}$ to the location of the relevant 
group of branes.

There is a third important geodesic: it also
does not move in the $z$-direction, but runs directly
from one group of $M$ D$3$-branes to the other.  The 
corresponding classical particle runs from one
peak to the other through the interconnecting saddle.   This
can be thought of as representing the massive $W$-particle
coming from the broken gauge symmetry, and lying in the
$(M,\bar M)$ or $(\bar M, M)$ representation depending upon 
the orientation of the geodesic.  The mass of this $W$ is given 
by \qenergy, but without the potential term (the constant of
integration for the $z$-motion is zero) and is thus given by
$m_W=\phi/\pi$.

Returning to the geodesics that correspond to a quark-antiquark
pair, we will see in the next section, that as we separate
the pair, {\it i.e.} increase $L$, the geodesic drops 
closer to the double-centered core.  Initially the pair is
in an unbroken phase, but at a critical separation there is
a phase transition and one can begin to distinguish the $(1,M)$ 
and the $(M, 1)$.  Associated with this transition, one also 
finds geodesics that correspond
to states with the quark, anti-quark and several
$(M,\bar M)$ or $(\bar M, M)$ $W$-particles.

\newsec{Analysis of Trajectories}

Using the potential derived in the preceeding section, we can now analyze
the quark-antiquark potential.   We first consider trajectories that 
come in at, or infinitessimally close to an angle of 90 degrees with 
respect to the $SU(M)\times SU(M)$ scalar axis.  The advantage of 
doing it this way is that the $\ZZ_2$ symmetry is not broken by the 
$U(1)$ scalar expectation value.  Moreover, the lowest trajectory
can be solved by quadratures.

\subsec{The geodesic on the symmetry axis}

Consider the trajectory that is aimed exactly at 
the saddle of the potential.  The trajectory then stays at 90 degrees 
since
there is no lateral force on the particle.  From the potential
in \doublepot, we see that we have an effective potential in $y$ only, 
while the $x$ coordinate stays fixed.  Using \fluxL\ and \qenergy, 
the length $L$ on the $D3$ brane  between the quarks is 
\eqn\zint{ L~=~2\int d\l ~ V(0,y(\l)) \ ,}
while the quark energy is given by the integral
\eqn\Eint{
E~=~{1\over2\pi}\int d\l \sqrt{(\partial_\l y)^2+2V(0,y(\l))} \ , }
where the term inside the square root is twice the conserved energy for
the particles motion.  (We have passed to the conventions
of \refs{\JMalda} by recaling $x \to \alpha' x$, $y \to \alpha' y$
and $\lambda  \to \alpha'^2 \lambda$ as described earlier.)

Using \conslaw\ and conservation of the particle energy,
 we see that as a function of $z$, $y$ satisfies
\eqn\yzeq{ 4V^2(0,y)(\partial_z y)^2=C-2V(0,y) \ ,}
where $C$ is an integration constant.  Hence, for a particle tractory
that starts at $y=\infty$ goes down to $y=y_0$ and then turns around
and goes back to $y=y_0$, we have
\eqn\zeq{\eqalign{
L&=\int_{y_0}^\infty {V(0,y) \over\sqrt{2V(0,y_0)-2V(0,y)} }dy\cr
&={2R^2(\phi^2+y_0^2)}\int_{y_0}^\infty {dy \over(y^2+\phi^2)
\sqrt{(y^2+\phi^2)^2-({y_0}^2+\phi^2)^2}} \ . }}
The above integral is a combination of elliptic integrals of the first and 
third kind. In terms of generalized  hypergeometric functions the 
length between  the quarks is
\eqn\zeqhyp{\eqalign{
L~=~{(2\pi)^{3/2}R^2\over (\Gamma(1/4))^2\sqrt{\phi^2+{y_0}^2}}
\Bigg[&\Fh{{\phi^4\over({y_0}^2+\phi^2)^2}}\cr
&+{\phi^2(\Gamma(1/4))^4\over48\pi^2(\phi^2+{y_0}^2)}~
\Gh{{\phi^4\over({y_0}^2+\phi^2)^2}}\Biggr]  \ . }}

The quark energy in \Eint\ is found using \conslaw\ and \yzeq.  Hence,
we find that
\eqn\energy{
E~=~{1\over\pi}\int_{y_0}^\infty {y^2+\phi^2\over
\sqrt{(y^2+\phi^2)^2-({y_0}^2+\phi^2)^2}}dy \ .}
This integral is divergent because the energy includes the quark bare
masses, which diverges as the position of the brane is taken to infinity.
We can find the finite potential piece by cutting off the integral at
$y_{max}$ and
subtracting off $y_{max}/\pi$, the sum of the quark masses.  Alternatively,
we can regularize the integral in \energy\ by replacing the numerator
of the integrand by $(\phi^2+y^2)^s$ and analytically continuing to
$s=1$.  In either case, the result is the same and the quark potential
energy is
\eqn\engam{\eqalign{
E~=~-~{\sqrt{2\pi}\sqrt{\phi^2+{y_0}^2}\over(\Gamma(1/4))^2}\Bigg[
&\Fg{{\phi^4\over({y_0}^2+\phi^2)^2}}\cr
&\qquad
-{(\Gamma(1/4))^4\phi^2\over16\pi^2(\phi^2+{y_0}^2)}~
\Gg{{\phi^4\over({y_0}^2+\phi^2)^2}}\Bigg] \ .}}

The asymptotic behavior of  $L$ for large and small $y_0$ is
\eqn\Lasym{\eqalign{
L&~=~{(2\pi)^{3/2}R^2\over(\Gamma(1/4))^2}~{y_0}^{-1}~+~{\rm O}
(1/y_0^3) \qquad\qquad y_0>>\phi\cr
L&~=~-{\sqrt{2}R^2\over\phi}~\log(y_0)~+~{\rm O}(1)
\qquad\qquad y_0<<\phi \ ,\cr}}
while the asymptotic behavior for the potential is
\eqn\Easym{\eqalign{
E&~=~-{\sqrt{2\pi}\over(\Gamma(1/4))^2}~y_0~+~{\rm O}(1/y_0)
\qquad\qquad y_0>>\phi\cr
E&~=~-{\phi\over\sqrt{2}\pi}~\log(y_0)~+~{\rm O}(1)
\qquad\qquad y_0<<\phi \ . }}
Hence we see that for small $L$ the quark potential has the 
coulombic behavior
\eqn\smallL{ E~\approx~-~ {4\pi^2\sqrt{4\pi g N}\over 
(\Gamma(1/4))^4}L^{-1} \ ,}
while for large $L$ the quark potential has the confining behavior
\eqn\largeL{
E~\approx~{1\over 2\pi\sqrt{4\pi gN}} \phi^2 L ~=~ {\pi \over 2}
{{m_W}^2 \over R^2} L  \ .}
The coulombic behavior is the same as that found for the unbroken
phase \refs{\RY,\JMaldb}.  This is not surprising; at short distances
the energy scale is much larger than the $W$ mass, so one expects
the potential to approach the coulomb potential for the unbroken case.

However, for large $L$ one would still expect to find a coulombic 
potential.  This is because when the energy scale drops well
below that of the $W$ mass, one would expect the $W$-particles
to become irrelevant, and so one should find the coulombic behavior
for an $SU(M)$ ($M = N/2$) super Yang-Mills theory.  Hence there 
should be other geodesics that describe this behavior.

\subsec{Geodesics around the symmetry axis}

In order to find such geodesics, we start by looking for geodesics 
that are very close to the straight trajectory.  Indeed, to lowest
order in perturbation theory we can assume that the motion in the 
$y$ direction is the same as for the straight trajectory, and take
the motion in the $x$ direction to be infinitesimally small.

The equation of motion in the $x$-direction is then
is approximately
\eqn\xmotion{
{d^2 x \over d \l^2}  ~=~
- {\partial^2 V(x,y)\over\partial x^2}\Bigg\vert_{x=0}~x ~=~
- {2R^4(5\phi^2-(y(\l))^2)\over(\phi^2+(y(\l))^2)^4}~x \ ,}
where $y(\l)$ corresponds to the motion of a straight trajectory.
Hence if $y(\l)^2>5\phi^2$ then the motion in the $x$ direction will
accelerate exponentially  away from the symmetry axis, $x=0$.  
However,  if  $y(\l)^2 < 5\phi^2$ then the there is a restoring force, 
and the motion is oscillatory.  Consider the geodesics that start at 
$x=0$ and $y=y_{max}\to\infty$ and return to the same point.
Such geodesics cannot exist unless they go close
enough to the origin for the $x$-motion to experience a restoring force
({\it i.e.} have $y(\l)^2 < 5\phi^2$).
It follows that if $L<<R^2/\phi$ then the only geodesic that starts
and finishes on the symmetry axis is the one that runs down the
symmetry axis and back.  There is thus a  phase of this theory
in which the $\ZZ_2$ symmetry is unbroken, and
the quark-antiquark pair feel only the full $SU(N)$
gauge theory.

Consider a geodesic whose initial velocity is such that $x$ is still
small when it enters the region with $y(\l)^2 < 5\phi^2$,
and begins to oscillate.  For the geodesic to begin and end on
the symmetry axis, $x=0$, the number
of half-period oscillations has to be an integer.  Hence, these modes will
be quantized.  We first examine the situation where there are a large
number of oscillations, which means that  many of the oscillations are
very close to the saddle.  We want to find the difference in
energy between the $n$ and $n+1$ half-periods.  The motions are almost
identical, except for an extra half-oscillation near the saddle.  For this
part of the motion $y$ can be approximated as a constant
$y \approx 0$, and so using \xmotion\ we find that the period for this
extra half oscillation is
\eqn\period{ \Delta\l~=~{\pi\over \sqrt{10}R^2}~\phi^{3}.}
Hence, using \fluxL, \doublepot\ and \largeL,
the increase in quark distance between the appearance of $n$ 
and $n+1$ oscillations is
\eqn\difflength{ \Delta L~=~\Delta\l R^4\phi^{-4}~=~{\pi R^2\over 
\sqrt{10}}~\phi^{-1} \ .}
and the change in the potential energy is
\eqn\diffenergy{\Delta E~=~{\pi\over 2\sqrt{10}}~{\phi\over\pi} \ .}
Thus, a trajectory with an extra half-oscillation appears when the energy 
is increased by an amount that is slightly less than half the $W$ mass,
$m_W = \phi/\pi$.

Once we have isolated a trajectory with a {\it fixed number}, $n$,
of small half-oscillations, we can
increase $L$ and find the asymptotic behavior for this trajectory.
As $L$ increases, the trajectory will move further and further up the
two peaks, or closer to the singularities of $V$, until it reaches 
the limit where the trajectory is the concatenation of a straight line
coming in from $(x,y)=(0,-\infty)$ to the peak at
 $(\phi,0)$ and then $n-1$  lines going back
and forth through the saddle between the peaks
and finally a straight line from one of the
peaks back to asymptotic infinity.  Figures 1 and 2 show the $L$ behavior
for geodesics with 2 and 6 half-oscillations.
Hence the total energy of this 
configuration is the sum of the quark masses plus $n-1$ times the $W$
mass $\phi/\pi$.  In other words, as $L\to\infty$, we find that we are left
with the two heavy quarks plus $n-1$ $W$-particles.
\goodbreak\midinsert
\centerline{\epsfysize2in\epsfbox{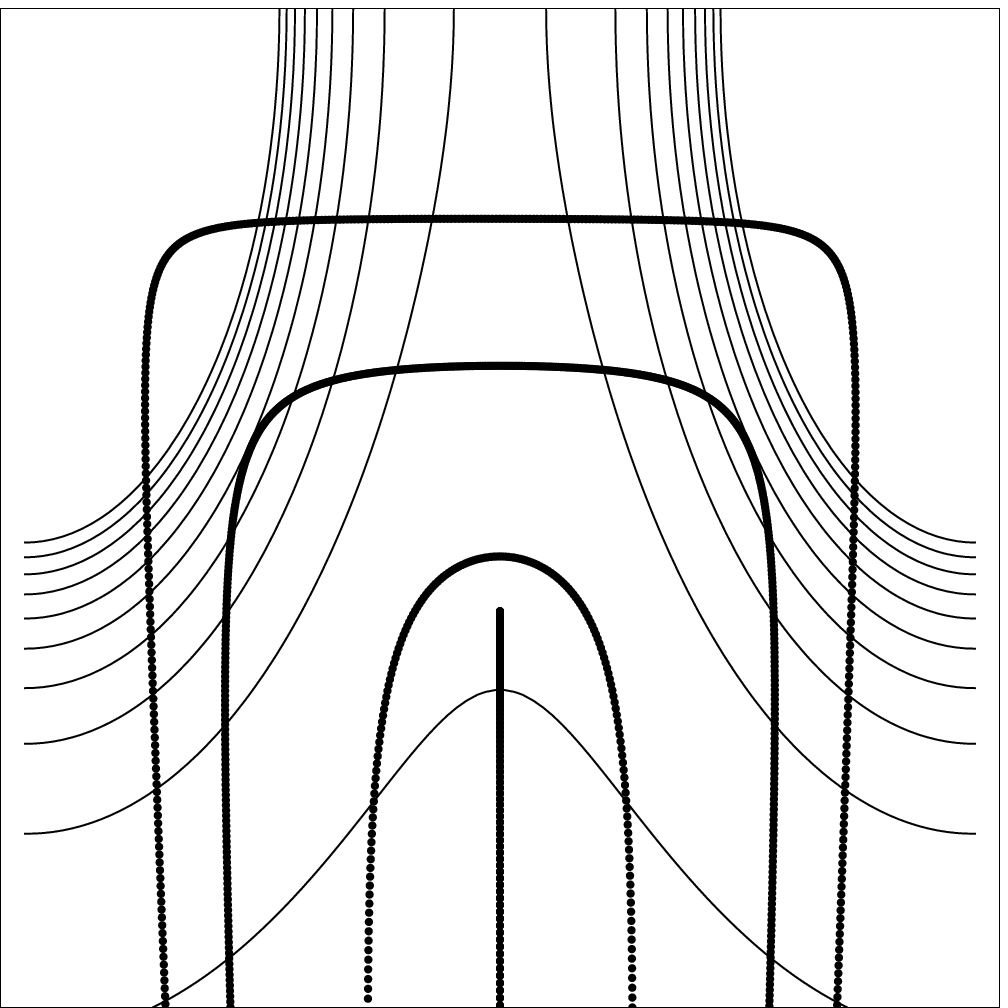}}
\leftskip 2pc
\rightskip 2pc\noindent{\ninepoint\sl \baselineskip=8pt {\bf Fig.~1}:
Geodesics with two half oscillations overlayed on a contour map of
the potential.  As $L$ is increased, the geodesic
approaches the concatenation of the BPS geodesics of two  quarks and a $W$.
The potential peaks are at the upper corners.  The straightline trajectory
in the figure is the small oscillation limit for two half-oscillations.}
\endinsert
\goodbreak\midinsert
\centerline{\epsfysize2in\epsfbox{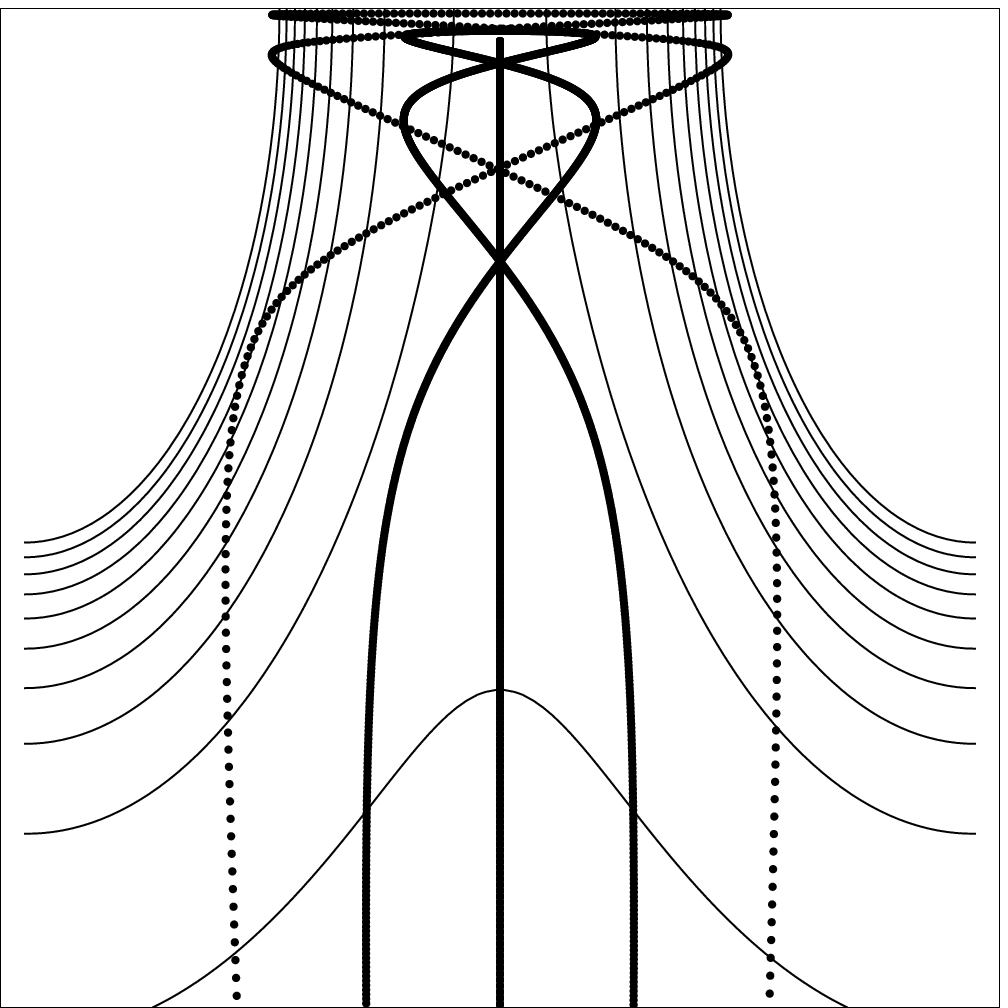}}
\leftskip 2pc
\rightskip 2pc\noindent{\ninepoint\sl \baselineskip=8pt {\bf Fig.~2}:
Geodesics with six half oscillations.  As $L$ is increased, the geodesic
approaches the concatenation of the BPS geodesics of two  quarks and a five
$W$s.}
\endinsert

It is also instructive to reverse this argument, and start with the 
heavy quarks separated by a large distance but with $n$ $W$'s in
between.   As $L$ becomes finite, the trajectory
moves off the peaks and has $n+1$ smooth half-oscillations. Hence this
corresponds to a situation with heavy quarks and $n$ intermediate 
$W$-particles.
 As $L$ is decreased
the oscillations become smaller and smaller, until we hit a critical 
distance $L_n$,
where the oscillations disappear all together and the trajectory
becomes our original straight-line trajectory.  As we decrease $L$ further,
the straightline trajectory becomes shorter until it coincides with
the trajectory with $n$ oscillations at $L=L_{n-1}$.  
Obviously, this trajectory corresponds to the case with heavy quarks and
$n-1$ $W$-particles.  Therefore, when $L=L_{n-1}$, the binding energy for
the last $W$ among $n$ $W$-particles is the $W$ mass $\phi/\pi$.

We can make a reasonable estimate of the strength of the coulombic potential
on the branch with $n$ $W$-particles.  
If $L$ is large, we can see from the figures
that the oscillations get very close to the peaks.  Moreover, each oscillation
reaches roughly the same distance $\e$ 
away from the peak with its velocity becoming quite small.  Hence we can
use the results of \refs{\RY,\JMaldb} to estimate $L$ and $E$.  For each
half-oscillation there is an ascent and a descent of a peak.  Hence, $L$
as a function of $\e$ is approximately
\eqn\Lncoul{
L~=~{2(\pi)^{3/2}(n+1)R^2\over(\Gamma(1/4))^2}~{\e}^{-1}~+~{\rm O}(1/\e^2),
}
where there is a factor of $n+1$ coming from the $n+1$ half-oscillations
and a factor of $1/\sqrt{2}$, since the effective gauge group at each peak
is $SU(N/2)$.  Likewise, the potential energy is
\eqn\Encoul{
E~=~-{\sqrt{2\pi}(n+1)\over(\Gamma(1/4))^2}~\e~+~{\rm O}(1) }
and hence $E$ as a function of $L$ is
\eqn\EnL{ E~\approx~-~ {4\pi^2(n+1)^2\sqrt{2\pi g N}\over 
(\Gamma(1/4))^4}~L^{-1} \ .}

This coulombic behavior can be understood as follows:  we can assume 
that there are $n$ intermediate $W$-particles equally spaced between 
the quarks. In the large $N$ limit each $W$ feels a coulombic 
attraction only with its two neighbors (the $W$-particles on the 
end feel an attraction to the neighboring quark), where the strength 
of this interaction is the coulombic quark anti-quark interaction 
for gauge group $SU(N/2)$ but for a separation of 
$L/(n+1)$.  Since there are $n+1$ interacting pairs, we expect the total
coulomb potential in \EnL.  Since the $W$-particles can move, we should
expect that these branches are unstable because of a coulomb instability.
We will discuss this further in the next subsection.

Figure 3 shows the branch of straight geodesics,  which becomes 
linear for large $L$, and the first three coulomb branches.  
The derivatives $dE/dL$ are continuous where the coulomb branches 
attach to the saddle branch, but $d^2E/dL^2$ is
not.  The lowest coulomb branch is the quark-antiquark branch with no 
intermediate
$W$-particles.  This branch is doubly degenerate, since a trajectory on this
branch
goes up one of the two peaks.  The second coulomb branch is a quark and
antiquark with an intermediate $W$-particle.  
There is only one branch since these trajectories
are $\ZZ_2$ symmetric.  Finally the third coulomb branch, which is also
doubly degenerate,  has two $W$-particles between 
the quark-antiquark pair.
\goodbreak\midinsert
\centerline{\epsfysize2in\epsfbox{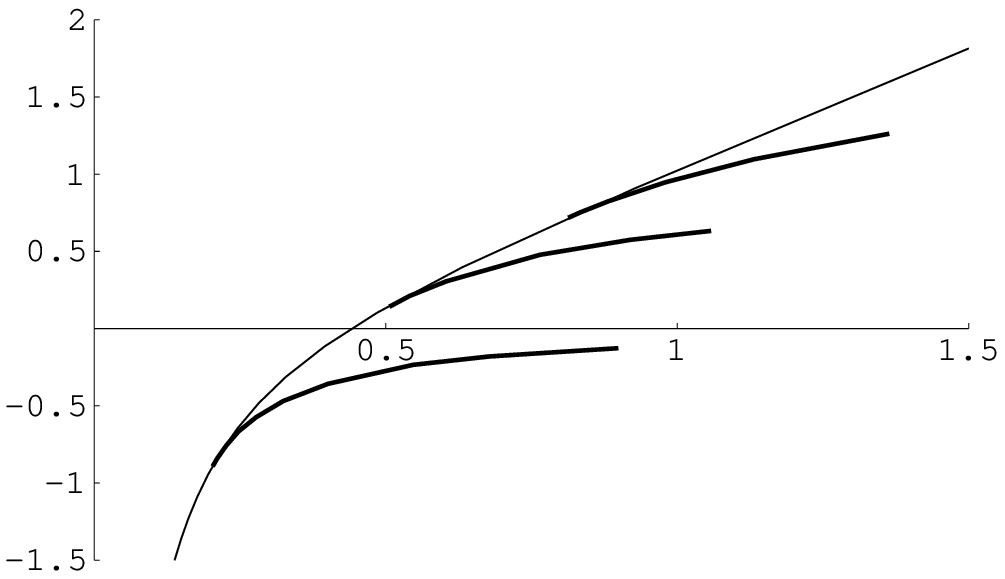}}
\leftskip 2pc
\rightskip 2pc\noindent{\ninepoint\sl \baselineskip=8pt {\bf Fig.~3}:
Plot of $E$ vs. $L$ for the saddle branch and the first three coulomb
branches.  $E$ is plotted in units of the $W$ mass, $m_W$, and $L$ is in units
of $R^2/m_W$.  The first and third 
branches are doubly degenerate.  The
second branch is a $\ZZ_2$ symmetric branch.}
\endinsert

It is instructive to compare the linear behavior of the straight trajectories
to the coulombic behavior of \EnL\ when $L=L_n$ and $n$ is large.  
In the linear regime \largeL\ yields
\eqn\dEdLlin{ {dE\over dL}~=~{\pi {m_W}^2\over 2R^2}~\approx~1.571 
{{m_W}^2\over R^2} \ .}
where $m_W = \phi/\pi$.  For large $n$ \difflength\ 
implies $L_n \approx {\pi R^2 n \over \sqrt{10} } \phi^{-1}$,
and hence the energy on the $n^{\rm th}$ Coulomb branch,
\EnL, yields
\eqn\dEdLcoul{ {dE\over dL}~\approx~{40\pi^2\over\sqrt{2} (\Gamma(1/4))^4}
{{m_W}^2 \over R^2}~ \approx~1.616{{m_W}^2\over R^2} \ .}
Hence, we see that the coulomb behavior of this branch persists all the
way down to where it merges with the linear branch.

We have thus seen that infinitesimal perturbations of
the symmetry axis geodesic yield geodesics corresponding to
a quark, anti-quark and $n$ intermediate $W$-particles experiencing
a net attractive coulombic potential.
We have also seen that there is a non-trivial threshold to
go from $n$ to $n+1$ intermediate $W$-particles.  This means that we should
view the symmetry axis geodesic as representing the process of
adiabatically separating the quark and anti-quark.  As the
separation gets larger, the number of $W$-particles that can be produced
between the quarks increases.
It is because of this increase in the $W$ number
that the energy does not undergo the Coulombic fall-off,
but instead exhibits the confining  behavior \largeL.

A small perturbation away from the straight geodesic leads 
to a selection of a definite $W$-number, and then further
adiabatic separation of the quark and anti-quark preserves this
number, but brings the state closer and closer to the
BPS threshhold.  It is easy to see that once a definite
particle number has been selected, the confining behavior of
$E$ must at least turn over, and revert to 
a Coulombic decay.  The reason is that
if the energy increased without bound, then {\it because the
particle number is fixed}, the energy will ultimately 
become far larger than the
total mass of all the $W$-particles between the quarks, and once
again scale invariance will effectively be restored.  
We thus have the following picture of $E(L)$ for all these
states:  the straight geodesic gives rise to and energy function
$E_*(L)$ that is akin to the Cornell potential, starting as 
$E_*(L) \sim - 1/L$, and
then crossing over to $E_*(L) \sim L$.  At finite, and ultimately
regular intervals, the graph of $E(L)$ versus $L$ bifurcates:
the confining potential, $E_*(L)$ continues, but a new
branch, $E_n(L)$, drops away and corresponds to a quark, anti-quark 
and $n$ $W$-particles.   This branch then  follows the Coulombic behavior 
\EnL, shifted by the mass, $n \phi/\pi$, of $n$ $W$-particles.  

\subsec{QCD Configurations and Stability}



As we discussed in section 3.2, the physical picture for the 
$(n+1)^{\rm th}$ coulomb
branch is that of $n$ $W$-particles equally spaced between the quarks.
This picture is confirmed by considering the energy
density, $\rho$, along the string, which is given by \endens. 
For a particle oscillating between the peaks, the QCD energy density
gets very small as the geodesics approach the peaks,
and becomes finite, or large as the geodesic passes across
the symmetry axis.  Indeed, as the particle moves higher and higher
up the peaks it spends most of its time on the peaks, and
very little time in between.   Thus the QCD energy density becomes 
very tightly peaked, and if the particle executes $n$ half-periods
then there are $n+1$ such peaks, the first and last being the
quark and anti-quark, and the remaining $n-1$ being $W$-particles.
For large $n$, the particle motion approaches the saddle and the 
oscillations  become regularly spaced and so the $W$-particles become
very regularly strung out along the QCD string between the quark 
and anti-quark.  Figure 4 shows plots of the string energy density 
for various values of $L$ on the coulomb branch with six intermediate 
$W$-particles. The plots show that the $W$-particles quickly localize 
as the quarks are pulled apart.
\goodbreak\midinsert
\centerline{\epsfysize1.2in\epsfbox{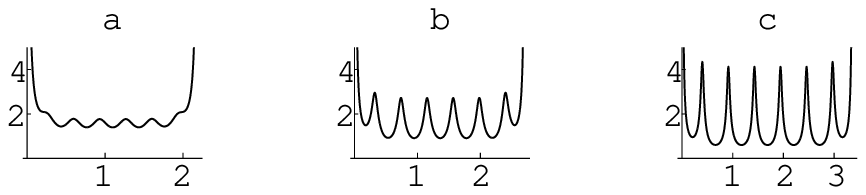}}
\leftskip 2pc
\rightskip 2pc\noindent{\ninepoint\sl \baselineskip=8pt {\bf Fig.~4}:
Plots of $\rho$ vs. $\ell$ for the coulomb branch with six intermediate $W$'s.
$\rho$ is in units of ${m_W}^2/R^2$ and $\ell$ is in units of $R^2/m_W$. 
The quark separations are (a) $L=2.19 R^2/m_W$ (b) $L=2.72R^2/m_W$ and
(c) $L=3.37R^2/m_W$.  The critical value for the appearance of this
branch is $L\approx 2.07 R^2/m_W$.}
\endinsert

Thus these oscillating geodesics correspond to $W$-particles
strung out like beads.  For the geodesic that runs up to the
saddle without oscillating, the $W$-particles are smeared out
along the length of the string, and as the oscillations 
increase the $W$-particles localize to a greater and greater
degree.  Because of this attractive Coulomb potential between the
beads on the string, we see that this configuration is very unstable,
and so we may infer that the linear flux tube is also unstable.

We can also see this instability of the flux tube using perturbative
analysis:  Consider geodesics that run straight up the saddle and look 
for small perturbations $\delta x$ in the $x$ direction.  Using 
equations \conslaw\ and \qenergy\ we see that perturbation of 
the energy is
\eqn\pertenergy{\eqalign{ \delta E&~=~{1\over4\pi}\int 
{dz\over\sqrt{\left({\partial y\over\partial z}\right)^2+H^{-1}}}
\left(
\left({\partial\delta x\over\partial z}\right)^2
-\ha H^{-2}{\partial^2 H\over \partial x^2}
\delta x^2\right)\cr
&~=~{1\over4\pi}\int d\lambda\left(
\left({\partial \delta x\over\partial\lambda}
\right)^2-{\partial^2 V\over\partial x^2}\delta x^2\right) \ .}}
The variation of the length is
\eqn\pertL{ \delta L~=~2\int d\l \left({\partial^2 V\over \partial x^2}
\delta x^2 +{\partial V\over \partial y}\delta y\right) }
and so a quadratic variation in $x$ can be compensated for by a linear 
variation
in $y$ in order that the length remain fixed.
Thus, we see that the string configuration is unstable if the operator
\eqn\linop{
{\cal L}~=~-{\partial ^2\over\partial\l^2}-{\partial^2 V\over\partial x^2}
}
has negative eigenvalue solutions.  But we know from \xmotion\
that a coulomb branch
occurs precisely when this operator has a zero eigenvalue.  Therefore,
every time the linear branch passes by a coulomb branch, an
eigenvalue of $\cL$ changes sign.  Hence if $L_n<L<L_{n+1}$ then the
linear branch has $n$ unstable modes.  In terms of $W$-particles, we
expect an unstable mode for each particle and so
the picture fits.

To summarize, we see that the linear branch is unstable because 
the flux tube is energetic enough to pop out intermediate 
$W$-particles between the quarks.  The appearance of
the $W$-particles is marked by a localization of the energy density
along the string between the quarks.  The presence of these
$W$-particles then destabilizes the configuration because of a coulomb
instability.  We have seen in \dEdLlin\ and \dEdLcoul\ that the coulomb
behavior of the $n$ branch persists all the way down to $L=L_n$, so we
expect very little softening of this instability on the linear branch.

When $L<L_1$, then the branch of straight trajectories is stable and is
in fact the only solution that extremizes the energy.  In this case
the $\ZZ_2$ symmetry is unbroken and the quark and anti-quark are in
the fundamental of $SU(N)$.   When $L>L_1$ then the stable branch
is the first coulomb branch which is doubly degenerate.  The degeneracy
of this branch reflects the fact that the quark and anti-quark are
fundamentals of one of the $SU(N/2)$ groups.  Thus, if we vary $L$
between values that are less than $L_1$ and greater than $L_1$ such
that we always remain on the stable branch, we see that there is a third
order transition in that $d^2E/dL^2$ is not continuous at $L=L_1$.  At
this point the $\ZZ_2$ symmetry is broken.

\newsec{Asymmetric and more general situations}

\subsec{Trajectories off the Symmetry Axis}

Up to this point we have been assuming that the $D3$-brane at infinity 
is on the symmetry axis defined by the two sets of $D3$-branes.  For
this configuration the heavy quarks are in the fundamental representation
for one of the two $SU(N/2)$ groups and both types of quarks have the
same BPS mass.  If the $D3$-brane at infinity is not on this axis, then the
quarks will have unequal masses.  If we denote the lighter quark by $q$
and the heavier quark by $q'$, then their mass difference is
$\Delta m=\phi \cos(\th)/\pi$, where $\th$ is the angle away from the
$D3$ brane axis.  So if $\th=0$ degrees then the difference in quark
mass is the mass of the $W$.

We now consider trajectories that are at an angle $\th$ at asymptotic
infinity.  Qualitatively, we expect to find behavior similar
to the symmetric case for large $L$.  That is we expect to find trajectories
corresponding to the coulomb potentials for  $q\overline q$ with some number 
$n$ of intermediate $W$-particles.  
As $L$ is decreased, the energy of this configuration
will match the potential energy for $q'\overline q$ with $n-1$ $W$-particles.  
Then
as $L$ is decreased even further this will match the potential energy for
$q'\overline q'$ with $n-2$  $W$-particles.  However, to make the next step to
$n-3$ $W$-particles, we have to first 
jump down to $q\overline q$ with $n-2$ $W$-particles, which 
releases a finite jump in energy because of the mass difference between
the $q$ and $q'$.  Hence these last two geodesics will not be continuously
connected.

Hence, we expect to find a lone geodesic that corresponds to  $q\overline q$
and which exists for any value of $L$.  All other geodesics will come in
bifurcating triple families.   That is, the $n^{\rm th}$ ``triple'' 
will correspond to $q\overline q$ with $n+2$ $W$-particles
 ($n$ is even), $q'\overline q$
with $n+1$ $W$-particles and $q'\overline q'$ and $n$ $W$-particles.  
These three families
of geodesics will be distinct for $L$ greater than some critical amount.  
At very large $L$, the strength of the Coulombic force will be 
different for the different branches, and the energy curves will 
have an asymptotic separation
of the $W$-boson mass.  As $L$ decreases the energy curves, $E(L)$,
must merge, as outlined above, and $E(L)$ must be singular (have a cusp) 
at the last merger.

One can easily see how this picture of triples emerges from
the symmetric picture considered in section 3.   Consider the functions, 
$E_*(L)$, and $E_n(L)$ described at the end of the last section.  For
$n$ even the relevant geodesics ultimately run up one peak or the
other, come to a halt and reverse their course, whereas for $n$ odd,
the geodesics never come to a halt and are symmetric between the peaks
as in Figures 1 and 2.  The energy functions $E_n(L)$, for $n$ even, are 
thus doubly degenerate, corresponding to $q\overline q$ and $n$ $W$'s,
or $q'\overline q'$ and $n$ $W$'s.  This degeneracy is lifted
when we split the degeneracy of the quark masses, and the bifurcating
curves $E_*(L)$ and $E_n(L)$ split into triples along the $E_n(L)$ 
curves for $n$ even.

Returning to the asymmetric situation, we can locate the critical values 
for the length at which mergers occur by examining the general behavior 
of the geodesics as a critical point is approached.  As we saw in the 
previous section, critical points may appear if a geodesic becomes unstable
due to a small oscillation.  A singularity of this type is shown in
Figure 5.  In this figure we will denote the quark associated with the
left-hand peak by $q$ and the other quark by $q'$.  We are considering 
geodesics with an incoming angle of 120 degrees, and there is one in particular
that rolls up the potential, stops, and then rolls back down.  This geodesic
is the limit of  geodesics that roll up one peak, move to the other
peak and then roll back down.  The latter class of geodesic may be thought
of as the state  $q'\overline qW$. The limit where this loop closes is
a singular point.   For this critical value of $L$, the $q'\overline qW$ 
trajectory merges into the $q\overline qWW$ trajectory.  
This second family of geodesics may be seen in upper part of Figure 6:
these geodesics roll up the left peak, roll over to the right, and then
reverse their course.  In terms of QCD,
the larger coulomb potential for $q\overline qWW$ compensates for 
the larger BPS mass of its constituents at the critical point.  For $L$
larger than this critical value one can be on either the $q\overline qWW$
or the $q'\overline qW$ family, while for $L$ less than the critical value 
these two families are merged and are indistinguishable. 

\goodbreak\midinsert
\centerline{\epsfysize2in\epsfbox{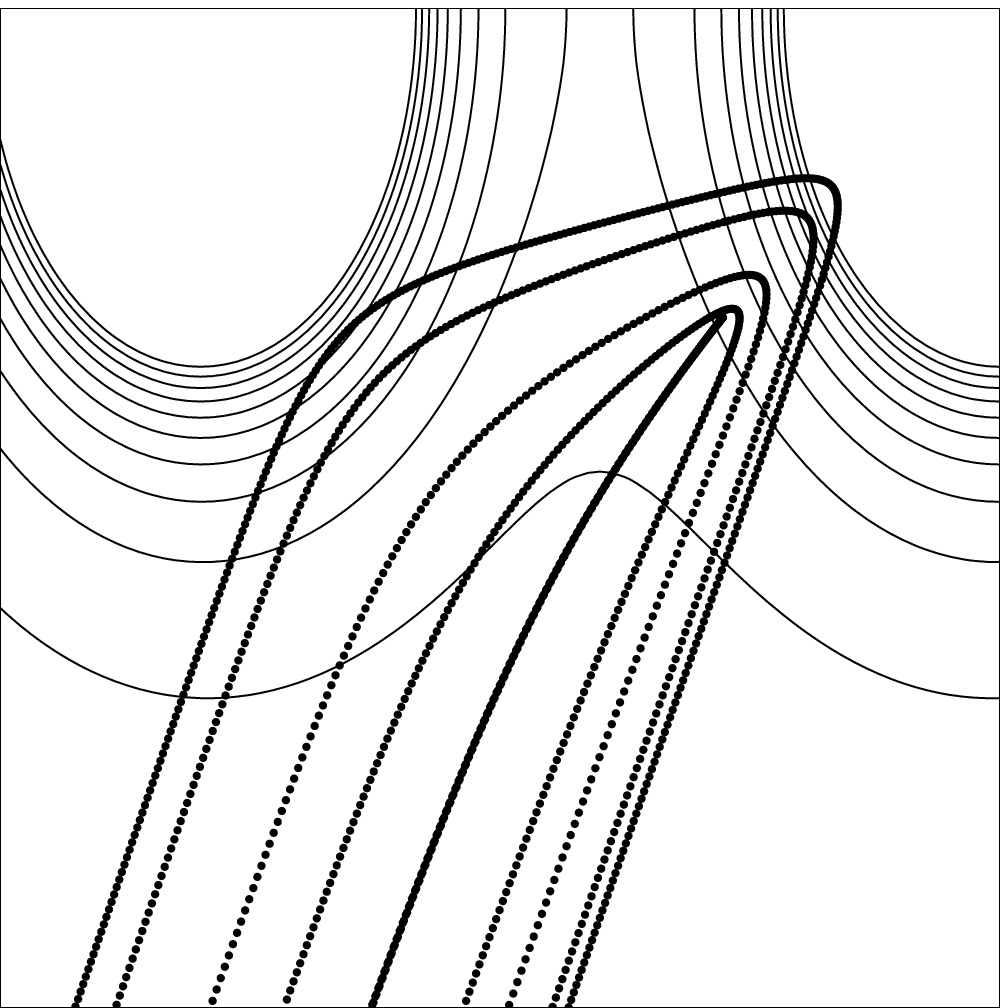}}
\leftskip 2pc
\rightskip 2pc\noindent{\ninepoint\sl \baselineskip=8pt {\bf Fig.~5}:
Geodesics with an incoming angle of 120 degrees that loop and cross
the symmetry axis twice.  For large $L$, the trajectory corresponds
to a $q'\overline q$ pair and an intermediate $W$.   There is a minimum value
of $L$ where the loop contracts to a line that retraces itself.  At this
value, the trajectory becomes identical with the trajectory for 
$q\overline q$ pair and two $W$-particles.}
\endinsert

There is another type of singularity which is not really a singularity
of the geodesics, but results in a singularity for the large $N$ Super 
Yang-Mills.  Suppose we have a family of geodesics parameterized by
some quantity $\s$.  Then $E$ and $L$ are both functions of $\s$.
However, $E$ and $L$ can be multivalued and if so there is some point $\s_0$
where $\partial_\s E=0$ and thus, by arguments given in section
2, $\partial_\s L=0$.  If we plot $E$ as a function of $L$ for these
families of geodesics, we would find a cusp where the $\s$ derivatives
vanish.  Hence this is a singularity and in fact corresponds to
the minimum $L$ and $E$ for these families of geodesics.  Figure 6 shows
a family of geodesics that have an incoming angle of 120 degrees and roll
up the second peak, stop and then roll back down.  As we discussed this
family contains a transition from  $q\overline q WW$ to  
a merger with $q' \overline q W$.  
Continuing beyond this we can smoothly deform these geodesics so that
the ``rest point'' of the particle moves down the peak and then moves back up 
it, finally representing a $q' \overline q'$ state.  Clearly this family
of geodesics will have a minimum value of $L$ since $L$  diverges as
the rest point of the particle moves up the peak.  At the minimum of
$L$, the $q'\overline q'$ trajectory merges with the $q'\overline q W$
trajectory.  Note that this critical point
does not necessarily occur when the particle energy is minimized.

\goodbreak\midinsert
\centerline{\epsfysize2in\epsfbox{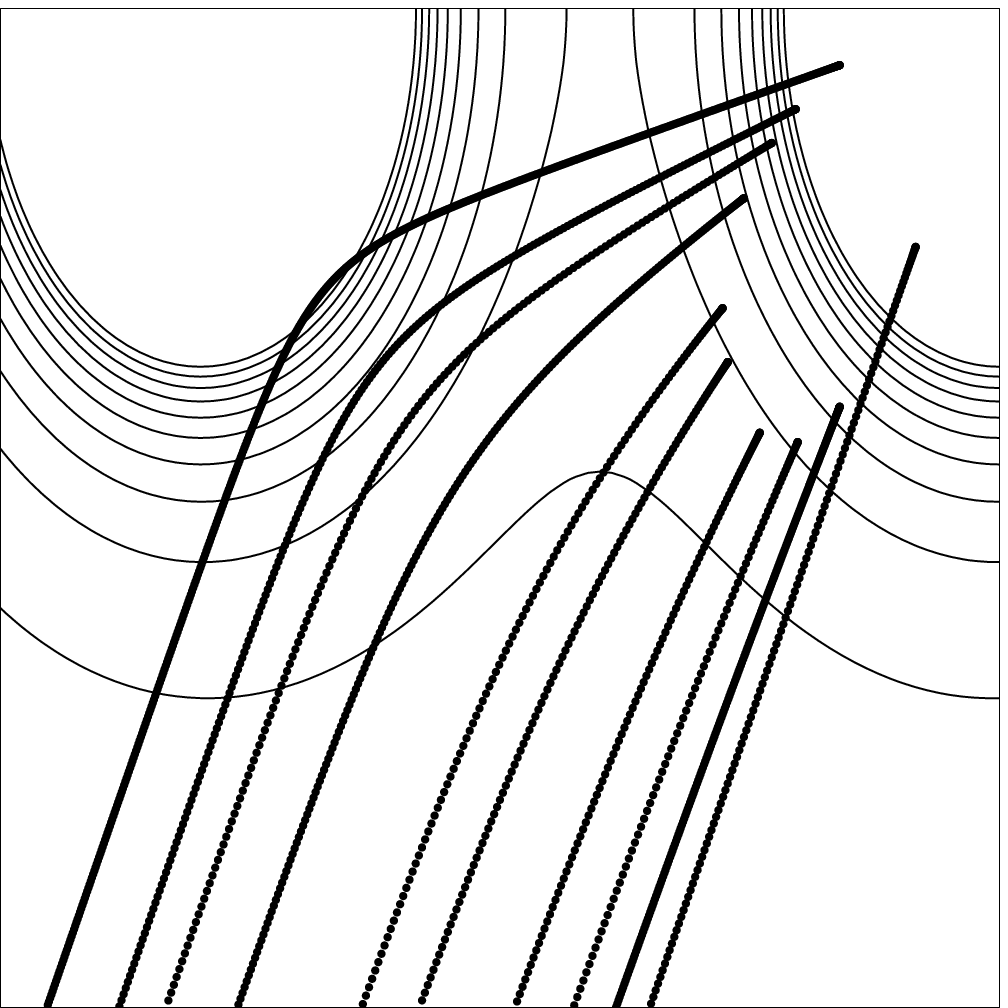}}
\leftskip 2pc
\rightskip 2pc\noindent{\ninepoint\sl \baselineskip=8pt {\bf Fig.~6}:
Geodesics with an incoming angle of 120 degrees that cross
the symmetry axis twice.  The trajectories
stop at some point and roll back in the opposite direction.  The figure
shows two branches for a given $L$.  For large $L$, one branch corresponds
to $q\overline qWW$ while the other branch
corresponds to a $q'\overline q'$ pair.  The fifth trajectory from the
left is the limit
of the loop trajectories in Figure 5.  The third trajectory from the right
corresponds to the trajectory with minimum length.}
\endinsert

We can also see in Figure 6 that there is a trajectory in this class of
geodesics where the particle energy $\mu$  is minimized.  Near this point, 
${dE\over dL}=1/\sqrt{2\mu}$ is roughly constant, so the potential is linear
in $L$.  Hence there is confining like behavior for trajectories in
the neighborhood of this point.  These trajectories are the merged  
$q'\overline q W$ trajectories.  This appears to
be a general feature.  That is, when a critical value $L_n$ is reached
such that the $q\overline q W^{2n}$ trajectories merge with the 
$q'\overline q W^{2n-1}$ trajectories, then for $L<L_n$ the potential is
roughly linear in $L$.  Figure 7 shows the lowest coulomb branch and the
first two triples.  In the lowest triple, $q\overline q WW$ merges with
$q'\overline q W$ which then merges with $q'\overline q'$.  Comparing
Figure 7 with Figure 3, one sees that the effect of the unequal mass
is to pull the degenerate branches in the symmetric case apart, leaving
isolated sets of branches.

\goodbreak\midinsert
\centerline{\epsfysize2in\epsfbox{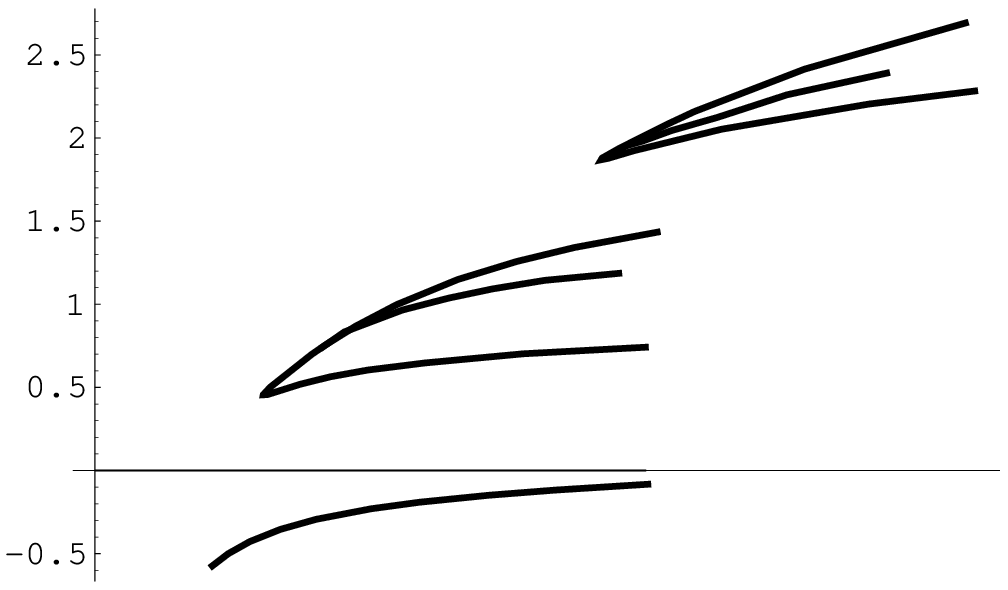}}
\leftskip 2pc
\rightskip 2pc\noindent{\ninepoint\sl \baselineskip=8pt {\bf Fig.~7}:
Plot of $E$ vs. $L$ for the $q\overline q$ branch, and the two lowest sets of
triples.  As $L$ is decreased the upper two branches in a triple merge
merge.  At even  smaller $L$ this branch merges with the bottom branch in
the triple.  To decrease $L$ even further requires a finite jump in energy
to a lower triple.
}
\endinsert

Thus once one moves off the symmetry axis, the interesting features
of the more symmetric situation survive, but in a more limited form.
For example, there still appears to be a nearly linear potential for
a range of separations, and in this range the number of $W$'s is
changing, but the asymmetry limits the amount of ``continuous'' change
in the particle number, and hence the behavior ultimately bifurcates into
various Coulombic limits.

The lone $q\overline q$ branch is obviously stable for any value of $L$
since it is the lowest energy state.  At $L$ less than some critical
value, $L_c$, this is the only available branch.  For $L>L_c$
 a bifurcated triple branch appears.  The lowest leg
of the triple is the $q'\overline q'$ branch.  While this branch has
more energy than the $q\overline q$ branch because of the higher quark mass,
this configuration is still a local minimum.  To see this, note that
in order for a $q'$ to switch to a $q$, we need to pop a $W$ near the quark,
but the $W$ will feel a net attractive force back to the quark.  If we
could pull the $W$ far enough away from the quark, it would eventually feel
an attractive force toward the $\overline q'$ antiquark, and then these 
two would combine to form a $\overline q$ antiquark.  Hence, we expect
the $q'\overline q'$ branch to be stable, but the branch destabilizes
when it merges with the $q'\overline q W$ branch, which occurs at the
cusp at $L=L_c$.  There is then a finite change in energy as the
configuration jumps down to the low branch.
 All other branches are unstable.

It is now relatively easy to see the general structure 
of what will happen in breaking $SU(N)$ to $SU(N_1) \times
SU(N_2)$.  First, there is still a saddle point, and thus
there will be a ``confining'' geodesic running up to it.
The confining geodesic will approach the two peaks
of the potential $V$ at some angle other than $90$
degrees.  One can view this angle as the asymmetry in the potential
being compensated for by an asymmetry in the quark masses.
There will be small oscillations about the confining geodesic,
and much of our earlier discussion will go through.  Indeed,
it is easy to convince onself that the asymmetric potential,
$V$, will result in families of geodesics that are deformations
of the ones considered earlier.  The corresponding 
conclusions about the quark-antiquark potential will
thus be completely analogous.

If one breaks $SU(N)$ to $p$ simple factors one can have a
much richer saddle-point structure in the potential $V$.  
This will lead to different ``confining geodesics,'' and
to far more complex structure.  In the $\ZZ_p$
symmetric breaking to $(SU(N/p))^p$, where the $D3$-branes
sit on the vertices of a regular $p$-gon, one might expect
that the $Z_p$ symmetric ``confining geodesic''  running along 
the symmetry axis of the $p$-gon would become more stable
if the curvature in the saddle is softened.

\newsec{Final Comments}

$N=4$ supersymmetric gauge theories are conformal and 
cannot confine, and yet we see geodesics that have 
some confining behavior.  There is no real conflict here 
since the ``confining geodesics'' 
are classically and quantum mechanically very unstable, 
and the corresponding QCD state will destabilize by popping out 
$W$-particles between the quarks, which are in turn
swallowed by the quarks because of the coulomb attraction.  In
the end we are left with two quarks feeling a coulombic force.
Of course linear flux tubes are not stable for real QCD either.  If
two quarks are pulled far enough apart, the flux tube destabilizes
by popping a shower of hadrons out of the vacuum.

On the more technical level, we have also found a simple relation 
for the interquark force in terms of a classical particle's energy.
Through this we have shown that the geodesic prescription of
\refs{\RY,\JMaldb} always leads to an attractive force.  It also gives 
a method for finding transitions in particle number, and for finding
confining families of geodesics.

\goodbreak
\vskip2.cm\centerline{\bf Acknowledgements}
\noindent

This work was supported in part
by funds provided by the DOE under grant number DE-FG03-84ER-40168.

\goodbreak

\bigskip

\listrefs
\end